\documentclass{article}
\usepackage{spconf,amsmath,graphicx}
\usepackage{multirow}
\usepackage{booktabs}
\usepackage{pgfplots}
\usepackage{subfig}
\newcommand\scalemath[2]{\scalebox{#1}{\mbox{\ensuremath{\displaystyle #2}}}}

\DeclareUnicodeCharacter{2212}{-}
\pgfplotsset{compat=1.7}
\pgfplotsset{
    every axis/.append style={
        scale only axis,
        width=0.17\textwidth,
        xtick={0, 0.25, 0.5, 0.75,1},
    },
    /tikz/every picture/.append style={
    }
}
\usepackage{mathtools}
\newcommand\Tstrut{\rule{0pt}{2.6ex}}         % = `top' strut
\newcommand\Bstrut{\rule[-0.9ex]{0pt}{0pt}}   % = `bottom' strut
\usepackage{color, colortbl}
\definecolor{LightCyan}{rgb}{0.88,1,1}

\title{Dynamic Graph Modeling of Simultaneous EEG and Eye-tracking Data For Reading Task Identification}
\name{Puneet Mathur, Trisha Mittal, Dinesh Manocha}
\address{University of Maryland, College Park}
\usepackage{color}
\begin{document}

\newcommand{\tm}[1]{\textcolor{magenta}{#1}}

%\ninept
%
\maketitle
\begin{abstract}

% Supervised machine learning techniques heavily relies on annotated data. Most of these annotations are acquired manually through crowd sourcing or through specialized human intelligence. However, there is no standard practice to quantify the quality of annotator efforts which is closely intertwined with the reading intent. Understanding the physiological aspects during the reading process can help differentiate normal reading and task-oriented reading during annotation for improving the manual labelling process as well as the quality of the annotations.

We present a new approach, that we call \textit{AdaGTCN}, for identifying human reader intent from Electroencephalogram~(EEG) and Eye movement~(EM) data in order to help differentiate between normal reading and task-oriented reading. Understanding the physiological aspects of the reading process~(the cognitive load and the reading intent) can help improve the quality of crowd-sourced annotated data. Our method, Adaptive Graph Temporal Convolution Network (AdaGTCN), uses an Adaptive Graph Learning Layer and Deep Neighborhood Graph Convolution Layer for identifying the reading activities using time-locked EEG sequences recorded during word-level eye-movement fixations. Adaptive Graph Learning Layer dynamically learns the spatial correlations between the EEG electrode signals while the Deep Neighborhood Graph Convolution Layer exploits temporal features from a dense graph neighborhood to establish the state of the art in reading task identification over other contemporary approaches. We compare our approach with several baselines to report an improvement of $6.29\%$ on the ZuCo $2.0$ dataset, along with extensive ablation experiments.

%  Co-registered Electroencephalogram (EEG) and Eye movement (EM) data during the reading activity can accurately quantify the cognitive load of the reader and aid in identifying the reading intent. Understanding the physiological aspects of the reading process~(the cognitive load and the reading intent) can help differentiate normal reading and task-oriented reading during annotation for improving the manual labelling process as well as the quality of the annotations. Towards this end, we propose Adaptive Graph Temporal Convolution Network (AdaGTCN), a neural architecture which introduces novel Adaptive Graph Learning Layer and Deep Neighborhood Graph Convolution Layer for identifying the reading activities using time-locked EEG sequences recorded during word-level eye-movement fixations. Adaptive Graph Learning Layer dynamically learns the spatial correlations between the EEG electrode signals while the Deep Neighborhood Graph Convolution Layer exploits temporal features from a dense graph neighborhood to establish the state of the art in reading task identification over other contemporary approaches.

\end{abstract}
\section{Introduction}

Reading is a complex cognitive task that requires the simultaneous processing of complex visual input across a series of brief fixation pauses and saccadic eye movements as well as retrieving, updating, and integrating contents of memory~\cite{Dimigen2014CoRegistrationOE, Kliegl2006TrackingTM, Radach2004TheoreticalPO}. Although each individual tends to process language in its own distinct style, the reading patterns tend to follow an underlying assumption that readers retain fixations on a word according to its cognitive importance in lexical processing~\cite{HorowitzKraus2018LongerFT}. Identifying such patterns in reading can help better model how humans read and perform regular linguistic tasks and transfer this knowledge to machines to better automate them for language processing. 

Understanding and automated modeling of human reading patterns can help in improving the manual crowd-sourced annotations in a variety of Natural Language Processing~(NLP) tasks as they are closely intertwined with the human subject reading intent \cite{hollenstein2019zuco}. While there is no standard practice to quantify the quality of annotator efforts, these annotations heavily feed into the NLP supervised learning setups effecting the quality of learning models.  Recognizing the reading patterns for estimating the reading effort also has broader applications in medical diagnosis of reading impairments such as dyslexia \cite{Syal2019TaskavoidantBA}, child developmental reading disorder \cite{Cavalli2017SpatiotemporalRO} and attention deficit disorder \cite{Kofler2019DoWM}.

% To further understanding in this domain, efforts have been made to learn from how humans perform these tasks. For instance, 

Hollenstein et al.~\cite{hollenstein2019zuco} defines two reading paradigms- Natural Sentence Reading~(NR) and Task-Specific Reading~(TSR). In a natural reading setup, the reader is expected to read sentences without any specific task other than comprehension. On the other hand, task-specific reading focuses on achieving some predecided linguistic goals beyond general comprehension such as relation extraction, sentiment labeling, pronoun resolution, and named entity recognition. In this work, we focus on identifying participant's reading activities - normal sentence reading vs task-specific reading using simultaneous psychophysiological signals - Electroencephalogram (EEG) and Eye Movement(EM).

The process of reading a specific text is considered complex from the perspective of neuroscience since it involves the vision, memory, motor control, learning, among others. Past research \cite{Dvok2018CognitiveBC} suggests the need of using brain Electroencephalography~(EEG) signals for studying the cognitive state of a subject for behavior analysis. Similarly, eye-tracking data also help better understand human reading patterns and is highly correlated with the cognitive load associated with different stages of text reading. Hence, simultaneous EEG and eye-tracking recordings hint towards promising results to further understand word-level brain activity signals during a reading session.  

\textbf{Main Contributions: }The following are some of the main contributions in this work. 
\begin{enumerate}
    \item We propose \textbf{A}daptive \textbf{G}raph \textbf{T}emporal \textbf{C}onvolution \textbf{N}etwork (AdaGTCN), a novel neural architecture composed of - Adaptive Graph Learning layer, Deep Neighborhood Graph Convolution layers and Dilated Inception Layers, all combined sequentially for identifying reading activities from synchronous EEG and eye-tracking data provided by ZuCo $2.0$ corpus \cite{hollenstein2019zuco}.
    \item Our proposed \textbf{Adaptive Graph Learning Layer} utilizes the spatio-temporal relationship between the EEG electrodes to dynamically learn the graph network structure based on EEG sequences during word-level eye-movement fixations and the \textbf{Deep Neighborhood Graph Convolution Layers} interleaved with \textbf{Dilated Inception Layers} simultaneously exploit the message passing in interdependent EEG electrodes while preserving their short range temporality.
\end{enumerate}

\section{Related Work}
\textbf{Understanding Human Reading Behavior: } Human reading analysis can directly benefit in determining the annotation complexity of text. \cite{Zheng2019HumanBI} conducted thorough investigation on the behavior patterns in complex reading comprehension to understand how humans allocate their attention during reading comprehension in an attempt to quantify reading efforts.

\textbf{Utilizing EEG and Eye-tracking Data: } Several previous works have explored EEG and eye-tracking features for analyzing human-information interaction tasks such as movie trailer analysis \cite{tauscher2017comparative}, diagnosis of mild Alzheimer's disease \cite{moghadami2020investigation} and hazard prediction \cite{kulke2016neural}. These studies concluded that isolated physiological signals miss out crucial information that are better perceived when combined together.  \cite{dimigen2011coregistration} studied the linguistic effects of co-registered eye movements and EEG neural activity in natural sentence reading and showed that such synchronized signals accurately represent lexical processing. This paper is the first attempt in utilizing synchronized EEG and eye-tracking signals for identifying forms of reading. Most prior works are handicapped by one or more of the following challenges when working with EEG and eye-tracking data either due to lack of requisite data or computational techniques. Our work can augment eye-tracking data for English-Chinese sight translation as done in \cite{Su2019IdentifyingTP}.

\textbf{Modeling EEG and Eye-tracking in Network Architectures: } Sequential modeling through recurrent neural networks like LSTM has been a popular technique for extracting relevant features from the EEG \cite{kuanar2018cognitive, Jang2018EegBasedVI} and eye-tracking data \cite{sims2019predicting, koochaki2019eye}. However, these models overlook the functional connectivities between the the EEG electrodes and their effect on the gaze patterns, leading to spatio-temporal information loss for sequence classification tasks \cite{zhang2019making}. We overcome this challenge by learning the interdependency between the EEG electrodes as part of the training process. Recently, many graph based methods have been proposed for extracting spatio-temporal information from EEG signals \cite{t1, jia2020sst, zhong2020eeg, zeng2020hierarchy}. However, most of these methods treat the graph network connections as static and do not learn the functional connectivities between the EEG electrodes as a part of the training process. We hypothesize that learning the graph structure in an online fashion helps the model exploit the latent spatial interdependencies in the brain signals without relying on external  biases about brain graph modeling.
\section{Problem Formulation}
We formally define the problem statement and the input and output data setup. For every sentence reading session, we have two signals, eye fixation data and the EEG signals. 

\noindent\textbf{Eye Fixation Data: }This data comprises of sequence horizontal axis gaze location entries for all individual fixations recorded while the reader fixates on the sequence of words ($w_{[0,n]}$) in a single sentence, where the $n$ words may not necessarily be in their natural linguistic order. \noindent\textbf{EEG Data: }The EEG signals have $p$ nodes (fixed number of electrodes corresponding to each frequency band). In a given sentence, the EEG signals recorded during the fixation duration $\Delta t_i$ for word $w_i$ are represented as $e^{g,i}_{p \times \Delta t_i}$ and its mean value across the time interval $\Delta t_i$ is given by $\hat{e}^{g,i}_{p}$. 

\noindent Given a participant reading a sentence in a reading session $S$ consisting of $n$ word fixations, the EEG time series is represented as $z =[\hat{e}^{g,0}_{p}, \hat{e}^{g,1}_{p}, \hat{e}^{g,2}_{p} \dots \hat{e}^{g,n}_{p}]$. We aim to learn a function $f(z) = y$ where $y=1$ if $S$ is Task-specific reading and $y=0$ for natural reading.

\section{Our Approach}
\begin{figure*}[h]
\centering
  \includegraphics[width=0.9\linewidth]{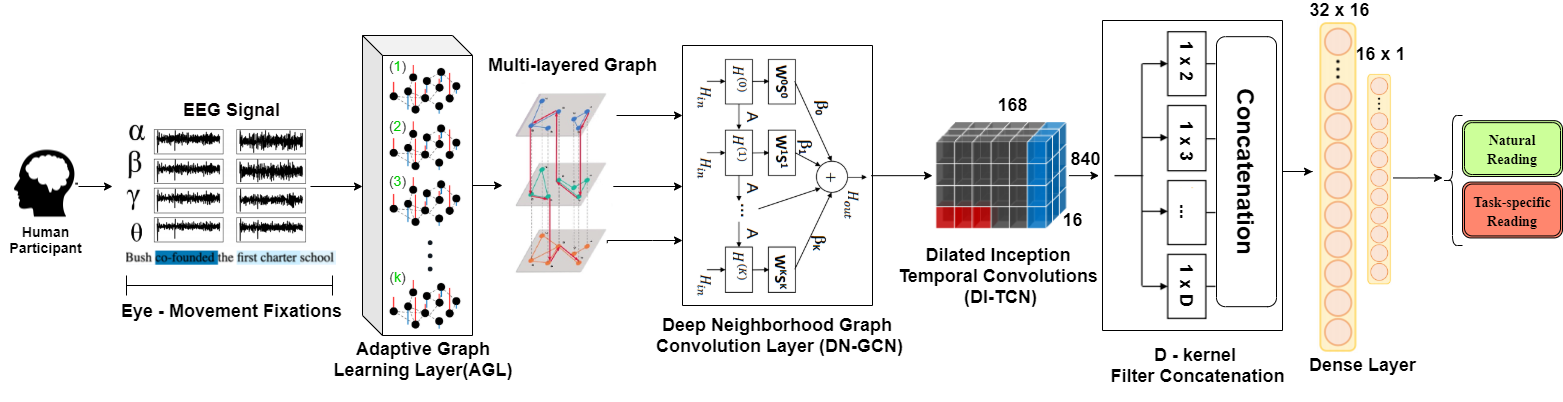}
  \caption{\small{Our network AdaGTCN consists of Adaptive Graph Learning (AGL) Layer which takes in the sequential time series input corresponding to each node and outputs a multilayered graph adjacency matrix. This is fed into the Deep Neighborhood Graph Convolution (DN-GCN) layer followed by the Dilated Inception Temporal Convolution (DI-TCN) layer. The DN-GCN module and the DI-TCN module both have 16 output channels. The output module consists of two dense layers having 32 and 16 output channels, respectively followed by a one neuron layer to normalize the output through Softmax to predict the class probabilities.}}
  \label{fig:architecture}
%   \vspace{-10pt}
\end{figure*}
In this section, we describe the individual components of the proposed Adaptive Graph Temporal Convolution Network (AdaGTCN) as illustrated in Figure \ref{fig:architecture}. 
%%%%%%%%%%%%%%%%%%%%%%%%%%%%%%%%%%%%%%%%%%%%%%%%%%%%%%%%
\subsection{Data Preprocessing}
EEG signals are long temporal sequences with long stretches of noisy artifacts. Thus, we    extract the EEG signals corresponding to the First Fixation Duration (FFD) - the duration of a fixation on the prevailing word, for each word in the reading sequence. Each electrode signal per channel is broken down in 8 frequency bands: $\theta_1$ (4–6 Hz), $\theta_2$ (6.5–8 Hz), $\alpha_1$ (8.5–10 Hz), $\alpha_2$ (10.5–13 Hz), $\beta_1$ (13.5–18 Hz), $\beta_2$ (18.5–30Hz), $\gamma_1$ (30.5–40 Hz), and $\gamma_2$ (40–49.5 Hz). Hence, for each word fixated by a participant in its reading sequence, we utilize the mean EEG signal values per channel in eight different frequency bands.
%%%%%%%%%%%%%%%%%%%%%%%%%%%%%%%%%%%%%%%%%%%%%%%%%%%%%%%%
\subsection{Adaptive Graph Learning~(AGL) Layer}
\label{sec:agl}
The pre-processed data is fed into an AGL layer that learns the adjacency matrix at the training time by randomly initializing node embeddings of multiple subsets of the graph and forms unweighted directed edges between the closest node pairs. Let a subset of nodes $v_i$ be randomly sampled from pool of all input nodes $V$. The input temporal node feature matrix $N$ of size $p \times n$ is passed through a linear layer with parameters $\theta_i$ with a $\tanh$ non-linearity. Additionally, each of the temporal node feature matrices is regulated by a saturation coefficient $\omega$ in order to provide a dropout effect. This ensures that the graph remains sparse and probability of dense hub formation reduces. The result of these operations is a sparse feature matrix $X_i$  as shown in Equation \ref{subgraph}.

\begin{equation}
\scalemath{0.9}{
    X_i = \tanh(\omega N \theta_i)
    }
    \label{subgraph}
\end{equation}

Repeating the process similarly for other $k$ node partitions we obtain $X_i, i \in \{0,1,2\dots k\}$. Each transformed sparse feature matrix $X_i$ is multiplied by every other $X_j$ where $j \neq i$ and summed together to form a partial adjacency matrix. Further, the sum of product of other sparse feature matrix with their transpose ($X_jX^{T}_j$) is subtracted  from the partial adjacency matrix through matrix regularization constant, $\lambda$ to regularize the complete adjacency matrix candidate $A_i$ by minimizing its diagonal component. The complete adjacency matrix  candidate is passed through a fully connected layer~($\phi$) with \textit{ReLu} non-linearity.
\begin{equation}
   A_i = \tanh(\sum^{k}_{j \neq i} (X_i X^{T}_j) - \lambda \sum^{k}_{j} X_j X^{T}_j)
\label{agg}
\end{equation}

\begin{equation}
   M_i =\text{ReLu}({\phi}(A_i))
\label{agg2}
\end{equation}

Equation \ref{agg} and \ref{agg2} is repeated for all $k$ node partitions. This stochastic sampling process assigns $k$ potential edge $e_{i,j}$ from vertex $u_i$ to $u_j$ for each vertex pair $(u_i, u_j)$. We select the top-k edge connections such that the top-k node candidates are set to 1 while the rest of the node pairs are set as zero. The sampled edge weights computed correspond to discrete variables. In such a case, the edge weights in the Adaptive Graph Learning Layer cannot be updated vie backpropogation. In order to convert these weights into continuous probabilistic distribution, we employ the Gumbel-softmax reparametrization technique \cite{jang2016categorical}, as given by $w_{i,j}= softmax((e_{i,j} + q)/\tau)$, to enable gradients computation. Here, $q$ is a random vector whose components are independent and identically distributed and $\tau$ is the softmax temperature that controls sampling smoothness. The process outlined in the above steps helps to learn stable node relationships over the training period. The hyper parameters $\omega$ and $\lambda$ are adjusted over the course of gradient backpropogation as new training data updates the model.
%%%%%%%%%%%%%%%%%%%%%%%%%%%%%%%%%%%%%%%%%%%%%%%%%%%%%%%%
\subsection{Learning Spatial Features in Neighborhood}

EEG electrodes connectivities attempt to simulate a miniature version of the human brain graph. The small hub assumption of the brain graph theory \cite{bassett2017small} describes the nerve connections in the human brain as a combination of non-random clustering with short path length. The effect of this assumption is critical in designing brain graph networks as a multi-layered graph structure that propagates information both in the immediate node locality and the deeper neighbourhood. However, past studies have shown that increasing the depth of the GCN converges it to the random walk’s limit distribution \cite{klicpera2018predict}. This inevitably results in information saturation where the hidden states converge to a single point and are skipped in the message passing in deeper layers. The traditional vanilla Graph Convolution (GCN) Layer \cite{kipf2016semi} is as $ H^{(K+1)}=\sigma \hat{A}H^{(K)}W^{(K)}$, where $H^{(K)}$ and $H^{(K+1)}$ are the input and output activation of the layer $K$, and $\hat{A} = D^{-\frac{1}{2}}(A+I_n)D^{-\frac{1}{2}}$ is a symmetrically normalized adjacency matrix with self-connections. $W^{(K)}$ is a trainable weight matrix such that $W^{(0)}$ is obtained as output from the AGL layer. The message passing algorithm in GCN acts as a simple neighborhood averaging operator which replaces each row in the feature matrix by the average of its neighbors. Inspired by \cite{abu2019mixhop}, we modify the vanilla GCN layer to introduce the Deep Neighbourhood Graph Convolution (DN-GCN) layer to recursively propagate the neighbourhood information from deep layers selectively over spatially dependent nodes. We modify the vanilla GCN layer with additional inputs from recursively deeper GCN layers to obtain the following equation: $\scalemath{.9}{H^{K+1}=\sigma(\sum^{K}_{l=1} \beta_l \hat{A} H^{l}W^l)}$, where $\beta_l$ is the depth regularization coefficient selected in a way such that $\sum^{K}_{l=1}\beta_l=1$. This modification retains a proportion of hidden states from each of the previous layers during the propagation step so as to preserve locality and exploring a deeper neighborhood at the same time. However, there can be cases when an EEG electrode node may not have any spatial dependency on its neighbours due to its peripheral location \cite{hollenstein2019zuco}. To handle such extreme cases, we add a feature selector term, $S^l$. For layers that do not have any spatial dependency on the $l^{th}$ layer, it is possible that $S^l=0$, except for $l=K$ to preserve information flowing through the self-connection acquired by the re-normalization trick of vanilla GCN layers. The final representation of the DN-GCN layer is given by $\scalemath{.9}{H^{K+1}=\sigma(\sum^{K}_{l=1} \beta_l \hat{A} H^{l}W^l S^l)}$.

% \begin{equation}
% \label{dngcn}
%     \scalemath{.8}{H^{K+1}=\sigma(\sum^{K}_{l=1} \beta_l \hat{A} H^{l}W^l)}
% \end{equation}

% \begin{equation}
% \label{finalgcn}
%      \scalemath{.8}{H^{K+1}=\sigma(\sum^{K}_{l=1} \beta_l \hat{A} H^{l}W^l S^l)}
% \end{equation}

\subsection{Extracting Temporal Patterns}

The EEG signals may have temporal dependencies spanning multiple ranges. Graph convolution layers can successfully model spatial dependencies but fall short of extracting time-varying patterns in long sequences. 1-D convolutional layers can be helpful in extracting temporal patterns. However, they are limited by their receptive field which grows in a linear progression with the depth of the network, requiring a deeper network that is harder to converge during optimization due to vanishing gradients. On the other hand, dilated convolutions \cite{yu2015multi} provide exponentially expanding receptive fields without losing resolution. Hence, the dilated convolution layers are a good choice for extracting temporal patterns of the input time series data in our model. However, choosing the right filter size is a difficult task as the temporal dependencies arising due to reading task difficulty have large variations. A very large filter size may miss critical recurring patterns while a shorter than required size may overfit the model. 
\begin{equation}
\label{conv}
     \scalemath{.9}{\Bar{x}_d = x \otimes f_{1 \times d} (t)= \int_{-\infty}^{\infty}f_{1 \times d}(s)x(t-r*s)}
\end{equation}

% \begin{equation}
% \label{DIL}
%     \Bar{x}=concat(\Bar{x}_1, \Bar{x}_2, \dots \Bar{x}_D)
% \end{equation}

Inspired by \cite{yang2019dilated}, we adopt Dilated Inception Network in our proposed architecture which uses multiple parallel dilated convolutions with different filter sizes to enrich the diversity of receptive fields in feature maps. We experiment various filters ($f_{1\times d}, d \in \{2,3,\dots D\}$) going till $D=12$. Given $x$ be the input 1-D sequence, the output of dilated convolution with dilation factor $r$, ($\Bar{x}_d$) obtained by convolution with filter $f_{1 \times d}$ is represented by Equation \ref{conv}. The transformed inputs across all dilation layers are truncated to the same length according to the largest filter and concatenated across the channel dimension to form the output of the DI-TCN layer as given by $\Bar{x}=concat(\Bar{x}_1, \Bar{x}_2, \dots \Bar{x}_D)$.
% shown in Equation \ref{DIL}.

\section{EXPERIMENTS and RESULTS}
\begin{table}[]
\centering
\resizebox{.95\columnwidth}{!}{
\begin{tabular}{ccccc}
\toprule
 && \textbf{Methods} & \textbf{F1}   & \textbf{Accuracy (\%)} \\
\midrule
&\multirow{6}{*}{\rotatebox{90}{Unimodal }}   &k-NN                           & 0.478          & 51.55                  \\
\multirow{8}{*}{\rotatebox{90}{\textbf{Baselines} }}      &&EEG-LSTM                        & 0.524          & 52.78                  \\
   & &EM-LSTM                          & 0.550          & 54.22                  \\
   & &EEG-GCN                       & 0.582          & 59.15                  \\
   & &EEG-GCN + Attention Pooling                  & 0.614         & 59.75                  \\
   & &EEG-GCN \Bstrut+ Hierarchial Pooling                    & 0.621          & 60.56                  \\ 
%   \multirow{2}{*}{\rotatebox{90}{Late Fusion}} &\textbf{Late Fusion Methods}\Tstrut& \textbf{F1}   & \textbf{Accuracy (\%)} \\
\cline{2-5}
& \multirow{2}{*}{\rotatebox{90}{\parbox{0.7cm}{Multi\\Modal}}} & EEG-LSTM + EM-LSTM\rule{0pt}{4.0ex}     & 0.640          & 62.33                  \\
&&EEG-GCN + EM-LSTM     & 0.659          & 63.50                  \\
\midrule

% &\textbf{Ablation }\Tstrut& \textbf{F1}   & \textbf{Accuracy (\%)} \\

 \multirow{4}{*}{\rotatebox{90}{\textbf{Ablation}}}\Tstrut
 &  &AdaGTCN                 & 0.695          &     69.79     \\
   & &AdaGTCN w/o DI-TCN                       & 0.652         &        64.12          \\
   & &AdaGTCN w/o DN-GCN                   & 0.633         &     63.72        \\
  &  &AdaGTCN w/o AGL                  & 0.675          &      66.20     \\
    \midrule
    \rowcolor{LightCyan}

&&\textbf{AdaGTCN~(Ours) }                      & \textbf{0.695} & \textbf{69.79}    \\

\bottomrule
\end{tabular}
}
\caption{\small{\textbf{Quantitative Results:} We compare with unimodal and multimodal baselines and perform ablation experiments and show an improvement of $6.29\%$}.}
\label{results}
% \vspace{-10pt}
\end{table}

\begin{figure}[h]
  \includegraphics[width=1\linewidth]{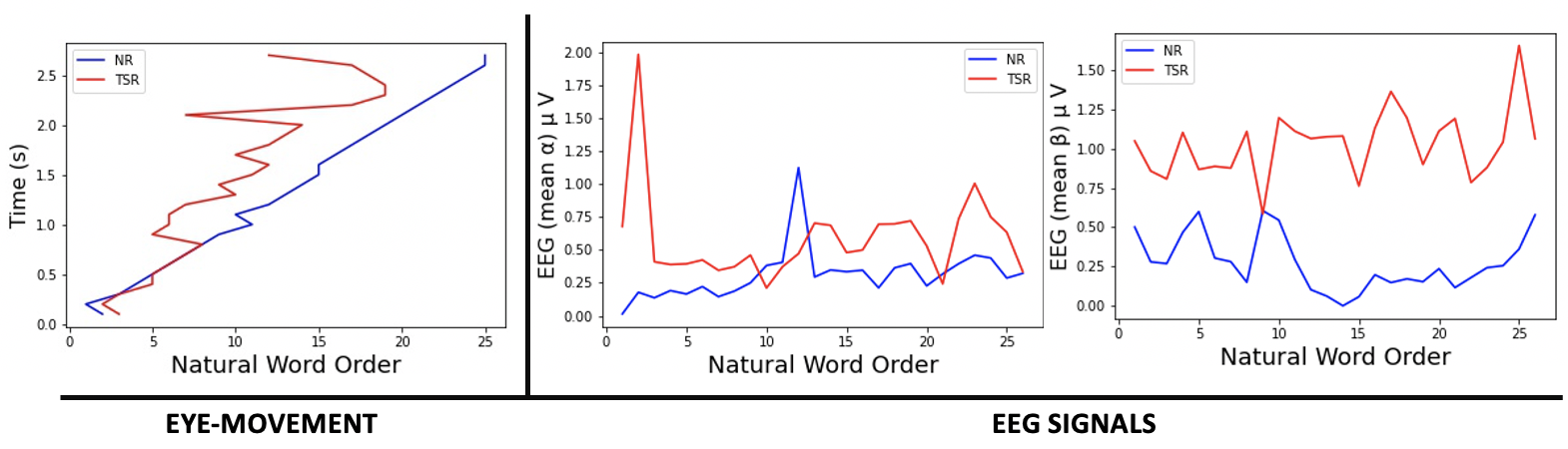}
  \caption{\small{\textbf{Qualitative Analysis:} \textit{Left:} Eye fixations on words with respect to time; \textit{Middle \& Right:} Mean EEG values across $\alpha$ and $\beta$ frequency bands for each word as they appear in the sentence. Comparison between natural reading (NR) and task-specific reading (TSR) for the sentence -``Henry Ford, with his son Edsel, founded the Ford Foundation in 1936 as a philanthropic organization with a goal to promote human welfare".}}
  \label{fig:case}
%   \vspace{-10pt}
\end{figure}

\noindent\textbf{Dataset: }ZuCo $2.0$ dataset \cite{hollenstein2019zuco} provides simultaneous EEG and eye-tracking data for $18$ participants reading a total of $739$ sentences~($349$ for NR and $390$ sentences for task-specific reading). The data of 12 participants ($66.67\%$) is used for train, $2$ for validation ($16.67$\%) and $4$ for testing ($33$\%). \textbf{Training Hyperparameters: } We summarize the range of hyperparameters as follows: dropout $\delta \in [0, 0.8]$, learning rate $\eta \in \{ 10^{−3}, 10^{−2}, 10^{−1} \}$, batch size $b \in \{4,8,16,32,64\}$ and epochs $(< 50)$. Adam was employed for optimizing the cross entropy loss of the model. The proposed model was trained on Pytorch using NVIDIA RTX 2080Ti, with batch size $4$ with the learning rate $0.01$ to converge in $20$ epochs. The input word-level fixation segmented EEG signal was padded to $168$, the max length of the input across the dataset. The hyperparameters of the Adaptive Graph Learning layer - saturation coefficient ($\omega$), matrix regularization constant ($\lambda$), and the temperature of Softmax smoothing ($\tau$) are all set to $0.5$ for best performance. The top-k edge connections in the same layer are experimented with the value of $k \in [1,20]$. Empirically, the depth of the node feature aggregation $K=2$ and the depth regularization coefficients ($\beta_1$, $\beta_2$) are set to $0.5$ and $0.6$, respectively in the DN-GCN layer. The filter sizes $f_{1xD}$ are experimentally found to be most effective when  $D\leq 7$. Layer Normalization is applied after each graph convolution module.

\noindent\textbf{Quantitative Results: } We compare with several baselines on the ZuCo 2.0 dataset and report the average values of F1 score and the accuracy, respectively, in Table~\ref{results}. Unimodal baselines include methods that process EEG and EM independently, either using recurrent neural networks such as LSTMs \cite{kuanar2018cognitive} or graph convolutional networks (GCNs) similar to \cite{t1}. In contrast, multimodal baselines perform late fusion to process both EEG as well as EM signals. Our proposed AdaGTCN model outperforms these baselines by a significant margin of \textbf{+6.29\%} in accuracy. To motivate the importance of the Adaptive Graph Learning Layer (AGL), Dilated Inception Temporal ConvNet layer (DI-TCN), and Deep Neighborhood GCN (DN-GCN) layer, we perform a series of ablation experiments, also shown in Table~\ref{results}. Our model generates sparser graphs with average node degree and total number of edges of $2.58$ and $1688$, respectively. This is a significant reduction in number of parameters in the best performing graph structure obtained by~\cite{Jang2018EegBasedVI} (average node degree=8 and total edges = 3524). \noindent\textbf{Qualitative Results: } We analyze the eye-movement across words and the EEG signals for one particular reading sequence in Figure~\ref{fig:case}. It can be observed that the eye-movement in natural reading is coherent with the reading sequence ordering as compared to the task-specific reading in the first plot. The subject tends to have multiple fixations and digressions on functional words such as names ("Henry Ford"), dates ("1936") and relations ("son"). The $\alpha$ and $\beta$ frequency range of EEG are relatively passive during natural reading, although sudden spikes can be observed in task-specific reading. 
% This is also verified in our experiments where the average node degree of $\alpha, \beta, \gamma$ and $\theta$ learnt by the model is 3.32 + 3.15 + 2.02 + 1.84, respectively. Lower average node degree reflects the fact that the node connectivity is stronger between the frequency ranges with more information flow and discriminate power for learning task specificty, {Trisha: Let's add this if space permits. Its a low-level detail, which is hard to follow anyway}

\section{Conclusion}
We propose Adaptive Graph Temporal Convolution Network (AdaGTCN), for identifying normal reading from task-specific annotation reading using simultaneous word-level eye-fixation segmented EEG signals. We motivate the advantages of learning the spatial graph structure formed by interdependent EEG electrodes at training time while exploiting the temporal patterns from a dense graph neighborhood. We demonstrate the benefits of AdaGTCN model through competitive performance on the ZuCo 2.0 dataset and benchmark relevant design choices for future signal processing applications on co-registered physiological data. Future direction of research will aim to leverage semi-supervised and unsupervised methods that do not rely on large amounts of annotataed data.

\section{Acknowledgement}

This work was supported in part by ARO Grants W911NF1910069 and W911NF1910315 and Adobe.

\bibliographystyle{IEEEbib.bst}
{\small\bibliography{refs}}

\end{document}

\newcommand{\tm}[1]{\textcolor{magenta}{#1}}

%\ninept
%
\maketitle

\section{Appendix}

\noindent\textbf{Dataset: }ZuCo 2.0 dataset \cite{hollenstein2019zuco} is the only known dataset for simultaneous Eye Movement (EM) and brain activity (EEG) recordings to analyze and compare normal reading to task-specific reading during annotation. It provides simultaneous time-locked EEG and eye-tracking data for 18 participants who read a total of 739 sentences - 349 sentences in a normal reading paradigm and 390 sentences in a task-specific reading paradigm (relation-annotation task) in a single sitting session under controlled experimental conditions. The data of 12 participants (66.67\%) is used for train, 2 for validation (16.67\%) and 4 for testing (33\%). This is done to ensure that models do not have access to any part pf a participants physiological or gaze patterns during the test time.
\newline

\noindent\textbf{Training Setup: } We summarize the range of hyperparameters as follows: dropout $ \delta \in [0, 0.8]$, learning rate $\eta \in \{10^{−3}, 10^{−2}, 10^{−1}\}$, batch size $b \in \{4,8,16,32,64\}$ and epochs $(< 50)$. Adam \cite{Kingma2015AdamAM} was employed for optimizing the cross entropy loss of the model. The proposed model was trained on Pytorch using NVIDIA RTX 2080Ti, with batch size 4 with the learning rate $1e-2$ to converge in 20 epochs. We repeat the experiment 10 times till convergence and report the average value of evaluation metrics - micro F1, accuracy, precision and recall. The hyperparameters of the AdaGTCN model were tuned on the validation set for minimal validation loss. The input word-level fixation segmented EEG signal was padded to 168, the max length of the input across the dataset. The hyperparameters of the Adaptive Graph Learning layer - saturation coefficient ($\omega$), matrix regularization constant ($\lambda$), and the temperature of Softmax smoothing ($\tau$) are all set to 0.5 for best performance. The top-k edge connections in the same layer are experimented with the value of $k \in [1,20]$. Empirically, the depth of the node feature aggregation $K=2$ and the depth regularization coefficients ($\beta_1$, $\beta_2$) are set to 0.5 and 0.6, respectively in the Deep Neighborhood GCN layer. The filter sizes $f_{1xD}$ are experimentally found to be most effective when  $D\leq 7$. Layer Normalization \cite{Ba2016LayerN}  is applied after each graph convolution module.

\bibliographystyle{IEEEbib.bst}
\bibliography{refs}